\RequirePackage[l2tabu,orthodox]{nag} 
\documentclass[12pt,letterpaper]{article} 

\usepackage[T1]{fontenc} 
\usepackage[utf8]{inputenc} 
\usepackage{ebgaramond} 
\usepackage{tgheros} 

\usepackage[USenglish]{babel} 
\usepackage[strict,autostyle]{csquotes} 
\usepackage[babel=true]{microtype} 

\usepackage{amsmath} 
\usepackage{authblk} 
\usepackage{booktabs} 
\usepackage{caption} 
\usepackage{datetime} 
\usepackage[final]{draftwatermark} 
\usepackage{endnotes} 
\usepackage{geometry} 
\usepackage{graphicx} 
\usepackage{natbib} 
\usepackage{rotating} 
\usepackage{setspace} 
\usepackage{titlesec} 
\usepackage{url} 

\usepackage{hyperref} 
\usepackage{orcidlink} 

\newdateformat{monthyeardate}{\monthname[\THEMONTH] \THEYEAR}

\newcommand{\myname}{Geoff Boeing}

\newcommand{\myorcid}{0000-0003-1851-6411}  
\newcommand{\myaffiliation}{Department of Urban Planning and Spatial Analysis\\University of Southern California}

\newcommand{\papertitle}{Modeling and Analyzing Urban Networks and Amenities with OSMnx}
\newcommand{\papercitation}{Boeing, G. 2025. \papertitle. \textit{Geographical Analysis}, published online ahead of print. \href{https://doi.org/10.1111/gean.70009}{doi:10.1111/gean.70009}}
\newcommand{\paperkeywords}{network science, openstreetmap, street networks, transportation, urban planning}

\graphicspath{{./figures/}}

\titleformat{\section}{\normalfont\sffamily\large\bfseries\color{black}}{\thesection.}{0.3em}{}
\titleformat{\subsection}{\normalfont\sffamily\small\bfseries\color{black}}{\thesubsection.}{0.3em}{}
\titleformat{\subsubsection}{\normalfont\sffamily\small\color{black}}{\thesubsubsection.}{0.3em}{}

\hypersetup{
    pdfauthor={\myname},
    pdftitle={\papertitle},
    pdfsubject={\papertitle},
    pdfkeywords={\paperkeywords},
    pdffitwindow=true, 
    breaklinks=true, 
    colorlinks=false, 
    pdfborder={0 0 0} 
}
\begin{document}

\title{\papertitle~\footnote{Citation info: \papercitation}}
\author[]{\myname~\orcidlink{\myorcid}}
\affil[]{\myaffiliation}
\date{}

\maketitle

\begin{abstract}

OSMnx is a Python package for downloading, modeling, analyzing, and visualizing urban networks and any other geospatial features from OpenStreetMap data. A large and growing body of literature uses it to conduct scientific studies across the disciplines of geography, urban planning, transport engineering, computer science, and others. The OSMnx project has recently developed and implemented many new features, modeling capabilities, and analytical methods. The package now encompasses substantially more functionality than was previously documented in the literature. This article introduces OSMnx's modern capabilities, usage, and design---in addition to the scientific theory and logic underlying them. It shares lessons learned in geospatial software development and reflects on open science's implications for urban modeling and analysis.

\end{abstract}

\section{Introduction}

Modeling networked urban infrastructure (such as streets, rails, canals, etc.) and geospatial features (such as amenities, buildings, transit stops, etc.) underpins analyses of travel behavior, accessibility, public health, sustainability, and spatial equity. This research spans the disciplines of urban morphology \citep[e.g.,][]{gervasoni_calculating_2017,dacci_signature_2019,coutrot_entropy_2022}, transport planning \citep[e.g.,][]{merchan_quantifying_2020,liao_disparities_2020,natera_orozco_data-driven_2020}, and network science \citep[e.g.,][]{feng_spatial_2020,yin_multi-task_2020,young_automatic_2020}. However, traditional limitations of data availability, inconsistent digitization standards, and a lack of well-documented, reusable tools historically limited the reproducibility, generalizability, scalability, and usefulness of empirical urban network science \citep{liu_generalized_2022}.

The open science movement has arisen in response to similar challenges across many disciplines. Open science seeks to make scientific processes and results transparent, interpretable, reproducible, and accessible to the public \citep{kedron_reproducibility_2021}. Open-source software is central to this movement. Developing, maintaining, and sharing well-documented, reusable, theory-infused analytical tools in an open commons prevents others from having to reinvent the wheel and accelerates broader scientific endeavors \citep{singleton_establishing_2016}. It also makes explicit the innumerable small modeling decisions and underlying assumptions encountered in all research \citep{gahegan_our_2018,rey_pysal_2022}. Geographers have long operated at the forefront of open science \citep{rey_show_2009,harris_more_2017}. Today, a robust open-source geospatial data science ecosystem exists to support the discipline's technical needs in representing and modeling geospatial data \citep{fleischmann_momepy_2019,rey_pysal:_2019,van_den_bossche_geopandasgeopandas_2024}.

Like their peers across disciplines, urban scholars need good tools for good science, including spatial network science. This article discusses OSMnx, a free, open-source, and fully type-annotated package written in pure Python that allows users to download, model, analyze, and visualize urban networks and geospatial features from OpenStreetMap data. Users can model walking, driving, biking, or customized networks with a single line of code and just as easily work with urban amenities and points of interest, building footprints, transit stops, elevation data, street orientations, speed and travel time, and routing. This package was first released in beta form in 2016 and that inchoate version was documented in a preliminary article \citep{boeing_osmnx:_2017}. At the time, it was the first tool enabling users to automatically download and model street network geometry and topology for any city in the world as a graph and then analyze and visualize it \citep{boeing_right_2020}. It has since become a widespread tool for a broad range of urban modeling and analysis in the intervening years \citep[e.g.,][]{coutrot_entropy_2022,dacci_signature_2019,feng_spatial_2020,gervasoni_calculating_2017,liao_disparities_2020,natera_orozco_data-driven_2020,yin_multi-task_2020,young_automatic_2020}.

However, since its initial release, the package has evolved drastically by expanding its functionality, improving performance, and stabilizing a user-friendly API.\@ To document these start-of-the-art capabilities and their scientific underpinnings in the literature, this article presents OSMnx's current modular structure, functionality, and applications for urban analytics. The following section introduces the theory, data, and tools of street network modeling and analysis, including OSMnx's place in that landscape. Then it presents the package's modular organization, capabilities, and usage. Finally this article concludes with a discussion of lessons learned for geospatial software development and the implications of open geographical science.

\section{Street Network Models and Analysis}

To describe precisely what this package does and how it works, we must first introduce some necessary theory underlying street network modeling. Innumerable scientific books and papers cover spatial network theory and analysis \citep[e.g.,][]{tinkler_graph_1979,barnes_graph_1983,gastner_spatial_2006,barthelemy_spatial_2011,ducruet_spatial_2014,fischer_spatial_2014,marshall_street_2018}. The following subsections discuss the fundamentals of spatial network models relevant to streets, and then briefly review the current landscape of analytics and tools.

\subsection{Model Fundamentals}

Real-world networks are commonly modeled as mathematical \textit{graphs} \citep{trudeau_introduction_1994, barthelemy_spatial_2022}. A graph, $G$, is a data structure consisting of two sets: one set, $N$, contains \textit{nodes} that are linked to each other by the second set, $E$, which contains node pairs (as sets or tuples) called \textit{edges}. Let $G = (N, E)$ and $\{u, v\} \subseteq N$ and $\{u, v\} \in E$. We can thus say for graph $G$ that: (1) edge $\{u, v\}$ links nodes $u$ and $v$, (2) edge $\{u, v\}$ is \textit{incident} to $u$ and to $v$, (3) $u$ is \textit{adjacent} to $v$ and vice versa, and (4) $u$ and $v$ are \textit{neighbors} \citep{newman_networks:_2010}. An adjacency matrix can fully represent a graph by defining these adjacent node pairs.

A node's \textit{degree} is the number of edges incident to that node. For example, a node with degree 2 has two incident edges which are consequently adjacent to each other. An edge can be \textit{undirected} (a 2-set, thus linking two nodes bidirectionally), \textit{directed} (a 2-tuple, thus linking one-way from a source node to a target node), or a \textit{self-loop} (linking one node back to itself). A graph with directed edges is called a directed graph, or \textit{digraph}. A graph that allows multiple edges to link a single pair of nodes is a \textit{multigraph} and those multiple edges are called \textit{parallel} edges.

A \textit{path} is a sequence of edges linking a sequence of nodes. The graph \textit{distance} between two nodes is the count of edges in the shortest such path between them. A \textit{weighted} graph's distance is the sum of some edge impedance attribute (e.g., length or time) along the shortest path minimizing that sum. A directed graph is strongly \textit{connected} if a path exists between each ordered pair of nodes, and it is weakly connected if such a path exists only if its edges are undirected. A disconnected graph contains multiple strongly or weakly connected \textit{components}---each a disjoint set of nodes forming its own internally connected subgraph.

A spatial network's nodes and/or edges are embedded in space \citep{barthelemy_spatial_2022}. Spatial graphs thus model both topology and geometry \citep{fischer_spatial_2014}.\ \textit{Topology} refers to the structure and configuration of the nodes and edges, whereas \textit{geometry} encompasses positions, lengths, angles, etc. A \textit{planar} graph's topology can be represented in a two-dimensional plane such that its edges intersect only at nodes \citep{barthelemy_modeling_2008}. Most street networks are nonplanar due to the occasional presence of overpasses and underpasses, but their spatial embedding and \enquote{approximate} planarity constrain their topological characteristics relative to other kinds of complex networks \citep{boeing_planarity_2020}. A \textit{primal} graph models a street network's intersections and dead-ends as nodes and its street segments as edges \citep{porta_network_2006-1}. A \textit{line} graph (sometimes called a \textit{dual} graph) inverts this topology to instead model streets as nodes and their intersections as edges \citep{porta_network_2006}, though it discards much of the network's geometry \citep{ratti_space_2004}.

\subsection{Street Network Analysis}\label{sec:street_network_analysis}

Graphs are commonly used for street network analysis, and researchers have applied many geometric measures of spatial networks to street network graphs \citep{barthelemy_spatial_2011}. Intersection density (i.e., the count of nodes with degree >1, normalized by network area) is perhaps the most common such measure of network \enquote{grain} in transport planning and urban design \citep[e.g.,][]{ewing_travel_2010}. The average street segment length (i.e., mean edge length) offers a linear proxy of block size. Street density is the sum of edge lengths normalized by network area. Circuity can take on different interpretations, including average circuity: the sum of edge lengths divided by the sum of great-circle distances between adjacent node pairs, representing the inverse of network edge straightness \citep{boeing_urban_2019}. Note that many such analyses should use undirected graphs to avoid double-counting bidirectional streets relative to one-ways.

Additionally, topological measures from graph theory are often applied to street networks \citep{barthelemy_spatial_2022}. The \textit{average node degree} (i.e., mean number of edges incident to the nodes) indicates graph connectedness\endnote{The separate term \textit{connectivity} has a distinct definition in graph theory, but is less useful for spatial network analysis because approximate planarity sharply constrains it: almost all real-world street networks have connectivity equal to 1.} and is perhaps the most common topological measure in transport planning and urban design \citep[e.g.,][]{barrington-leigh_century_2015,barrington-leigh_more_2017,barrington-leigh_global_2020}. Networks with high connectedness can be more robust against perturbation as they offer alternate routing options if parts of the network fail \citep{boeing_resilient_2024}. Various measures of \textit{centrality} are also common \citep{crucitti_centrality_2006}. For example, a node's \textit{betweenness centrality} measures the share of all graph shortest paths that pass through the node \citep{barthelemy_betweenness_2004,barthelemy_self-organization_2013}. The graph's maximum betweenness centrality indicates the share of shortest paths that rely on its most important node: high values suggest possible chokepoints that represent single points of failure, such as a bridge connecting a city's halves across a river \citep{boeing_resilient_2024}.

In street network analysis, researchers use these geometric and topological measures to characterize a network's form. Such analyses often also employ path solving \citep{miller_measuring_1999,wang_road_2020}. For example, accessibility analyses solve shortest paths (often weighted by length or travel time) from origin nodes (e.g., homes) to destination nodes representing the locations of amenities (e.g., workplaces, schools, transit stops, parks, grocers, hospitals, etc.) to measure access \citep{foti_behavioral_2014,liu_generalized_2022}. Disaster analyses often simulate emergency responses or evacuations along the network to understand traffic flows following a disaster \citep{sasabe_road_2020,tamakloe_finding_2021}. These kinds of analyses can reveal different outcomes for different communities or locations within the city, offering guidance to target practitioners' interventions.

\subsection{Street Network Tools and Data}

Various tools exist to model and analyze spatial networks like street networks. For example, ESRI's ArcGIS software includes a Network Analyst extension and QGIS offers plug-ins for network analysis. However, such GIS tools' network data ingestion and analysis capabilities are usually fairly limited. Conversely, dedicated network analysis tools such as Gephi, IGraph, and NetworkX offer robust network analysis functionality but limited or no geospatial capabilities. Other dedicated spatial network tools exist, often for specific analytical purposes, including PySAL spaghetti for network inference \citep{gaboardi_spaghetti_2021,rey_pysal_2022}, Pandana for urban accessibility \citep{foti_behavioral_2014}, momepy for urban morphology \citep{fleischmann_momepy_2019}, and stplanr for transport planning \citep{lovelace_stplanr_2019}.

Street network data can come from various sources, including governmental sources like the US Census Bureau's TIGER/Line ShapeFiles which represent network geometry but lack sufficient topological information to build a properly nonplanar model \citep{boeing_graph_2025}. Alternatively, OpenStreetMap offers a worldwide, public web mapping platform and geospatial database that anyone contribute to, with some editorial oversight, and includes streets, highways, transit, buildings, footpaths, cycleways, points of interest, and political boundaries \citep{jokar_arsanjani_openstreetmap_2015}. Although its coverage varies, it includes network geometry and topology and its data quality is generally high---particularly so in urban areas, but with notable exceptions in parts of China and sub-Saharan Africa \citep{barron_comprehensive_2014,barrington-leigh_worlds_2017}.

OpenStreetMap's data model comprises three \textit{element} types: \textit{nodes} (points), \textit{ways} (i.e., sequences of nodes, which can be either open ways representing lines or closed ways representing polygons), and \textit{relations} (i.e., between nodes and/or ways). These elements can possess one or more \textit{tags}: key-value pairs containing attribute data. Geometry is defined by nodes' latitude and longitude coordinates and OpenStreetMap provides one-dimensional node-edge topology. However, it does not provide this neatly in the graph theory sense. Rather, every OpenStreetMap way element is formally composed of an $n$-tuple of OpenStreetMap node elements, plus attribute data. Yet these relationships are topological: they are preserved under continuous deformations and transformations of the underlying space.

OpenStreetMap data are available for download from third-party services, such as Geofabrik, and web APIs including Overpass and Nominatim. Historically, however, constructing a street network model from raw OpenStreetMap data was fairly difficult. For example, the Overpass API can be cumbersome to extract the right data to construct an appropriate spatial graph. Analysts usually had to write extensive ad hoc code to process the raw data into a useful model and then write more code to conduct analyses. Their decisions on how to handle ubiquitous street network characteristics---such as directedness, planarity, self-loops, parallel edges, or culs-de-sac---frequently went undocumented or even unconsidered. With every research team writing its own ad hoc code, innumerable small modeling decisions inevitably would go unreported in the resulting peer-reviewed literature---but these small decisions add up to inhibit interpretability and replicability.

\section{The OSMnx Package}

\subsection{Overview}

OSMnx (pronounced as the initialism: \enquote{oh-ess-em-en-ex}) primarily fits into this landscape by enabling users to build a graph model of a spatial network anywhere in the world by automatically downloading raw data from OpenStreetMap and constructing a spatialized NetworkX model. This allows users to avoid unnecessary ad hoc coding and replaces myriad small modeling decisions with a well-documented, transparent, common tool.

In turn, OSMnx has recently become a standard tool in the literature for both retrieving urban data broadly and for modeling street networks specifically. For example, it has been used by researchers, students, and practitioners across a wide variety of fields, including transport planning \citep[e.g.,][]{natera_orozco_data-driven_2020,dumedah_case_2020}, vehicle-sharing systems \citep[e.g.,][]{luo_d3p_2020,zhang_electric_2019}, urban centrality modeling \citep[e.g.,][]{wang_road_2020,torres_alisis_2019}, computer science \citep[e.g.,][]{yin_multi-task_2020,young_automatic_2020}, persistent homology \citep[e.g.,][]{feng_spatial_2020,abdelkader_topological_2018}, synthetic population modeling \citep[e.g.,][]{tuccillo_likeness_2022,tuccillo_spatial_2023}, calculating urban indicators \citep[e.g.,][]{quistberg_building_2019,wang_impacts_2020}, and trip simulation \citep[e.g.,][]{hernandez-hernandez_anger_2019,merchan_quantifying_2020}.

\subsection{Modular Organization}\label{sec:modular_organization}

Here we provide an up-to-date\endnote{This article documents OSMnx version 2.0.1, released January 2025.} description of the package's modular organization, capabilities, and usage. OSMnx is structured as a collection of 17 modules that organize related functionality (Table~\ref{tab:osmnx_modules}). It uses the data structures of NetworkX \citep{hagberg_exploring_2008}, a network analysis Python package, and GeoPandas \citep{van_den_bossche_geopandasgeopandas_2024}, a Python package for working with geospatial dataframes. It interacts with three public web APIs to collect data: the OpenStreetMap Nominatim API, the Overpass API, and the Google Maps Elevation API (or their equivalents with the same interface).

\begin{table*}[p!]
    \centering
    \caption{The OSMnx v2.0 public API's modules and the functionality they expose.}\label{tab:osmnx_modules}
    \begin{tabular}{lp{9.7cm}}
        \toprule
        Module &  Functionality \\
        \midrule
        \texttt{bearing}  & Calculate graph edge compass bearings and orientation entropy.\\
        \texttt{convert}  & Convert a graph to/from different data structures.\\
        \texttt{distance}  & Calculate spatial distances and find nearest graph nodes or edges to points.\\
        \texttt{elevation}  &  Attach node elevations from raster files or a Google Maps compatible elevation API, and calculate edge grades.\\
        \texttt{features}  &  Download OSM geospatial features' geometries and attributes, such as points of interest, building footprints, transit stops, etc.\\
        \texttt{geocoder}  & Geocode place names or addresses or retrieve OSM elements by place name or ID, via the Nominatim API.\\
        \texttt{graph}  & Download and create graphs from OSM data, using filters to query the Overpass API for built-in network types or a custom filter.\\
        \texttt{io}  & Save/load graphs to/from GraphML, GeoPackage, or OSM XML files.\\
        \texttt{plot} & Visualize street networks, routes, orientations, and geospatial features.  \\
        \texttt{projection}  &  Project a spatial graph to a different coordinate reference system.\\
        \texttt{routing}  &  Calculate graph edge speeds, travel times, and weighted shortest paths between nodes.\\
        \texttt{settings}  & Configure global package settings.\\
        \texttt{simplification}  & Simplify and consolidate spatial graph nodes and edges.\\
        \texttt{stats}  &  Calculate geometric and topological network measures.\\
        \texttt{truncate}  & Truncate a spatial graph by distance, bounding box, or polygon.\\
        \texttt{utils}  & General utility functions.\\
        \texttt{utils\_geo}  & Miscellaneous geospatial utility functions.\\
        \bottomrule
    \end{tabular}
\end{table*}

\subsubsection{Geocoding and Querying}

OSMnx geocodes place names and addresses via the Nominatim API\@. Users can use the \texttt{geocoder} module to geocode place names or addresses to latitude-longitude point coordinates. Or, they can retrieve place boundaries or any other OpenStreetMap elements by name or ID.\@ Using the \texttt{features} and \texttt{graph} modules, as described below, users can download OpenStreetMap data from the Overpass API by latitude-longitude point coordinates, address, bounding box, bounding polygon/multipolygon, or place name (e.g., neighborhood, city, county, etc.).

\subsubsection{Urban Amenities}

Using OSMnx's \texttt{features} module, users can search for and download any geospatial features (e.g., points of interest, building footprints, grocery stores, schools, public parks, transit stops, etc.) from the OpenStreetMap Overpass API as a GeoPandas \texttt{GeoDataFrame} object. This uses OpenStreetMap tags to search for matching elements within the prescribed query area.

\subsubsection{Modeling Networks}

Using OSMnx's \texttt{graph} module, users can download any spatial network data (such as streets, paths, rail, canals, power lines, etc.) from the Overpass API and model them as a graph. This process uses filters to query the Overpass API:\@ users can either specify a built-in network type (e.g., bike, drive, walk, all, etc.) to retrieve matching elements, or they can provide their own custom filter written in OverpassQL.\@

OSMnx models spatial networks as primal, nonplanar, weighted, directed multigraphs with possible self-loops---specifically, these are NetworkX \texttt{MultiDiGraph} data structures \citep{hagberg_exploring_2008}. It models a one-way street as a single directed edge from node $u$ to node $v$, but a bidirectional street is modeled with two reciprocal directed edges (with identical geometries)---one from $u$ to $v$ and another from $v$ to $u$---to represent both possible directions of flow. Because these graphs are nonplanar, they correctly model the topology of interchanges, overpasses, and underpasses. That is, edge crossings in a two-dimensional plane are not nodes in an OSMnx model unless they represent true junctions in the three-dimensional real world.

Under the hood, OSMnx does several things to produce the best possible model, many of which aim to attenuate the \enquote{artificial perimeter} problem of making a graph model of a real-world network. It initially buffers the query boundary by 500 meters to create a preliminary graph before truncating it to the user's desired query boundary: this ensures accurate streets-per-node counts around the perimeter. As the graph may contain some disconnected components, OSMnx returns the largest weakly connected component by default, though this is configurable. This parameterization allows users to control which trade-off they prefer: either (1) retaining everything within the boundary even if it includes some small disconnected components that inhibit routing, or (2) discarding some elements within the boundary to ensure graph connectivity. OSMnx also automatically simplifies the graph topology as discussed below.

\subsubsection{Graph Simplification}

The \texttt{simplification} module automatically processes the network's topology from the original raw OpenStreetMap data to ensure that individual nodes represent individual intersections or dead-ends and edges represent the street segments that link them. This graph simplification is of two primary types: edge simplification and node consolidation. These algorithms are described in detail in \citet{boeing_graph_2025} and summarized here.

\begin{figure*}[tbp]
    \centering
    \includegraphics[width=1\textwidth]{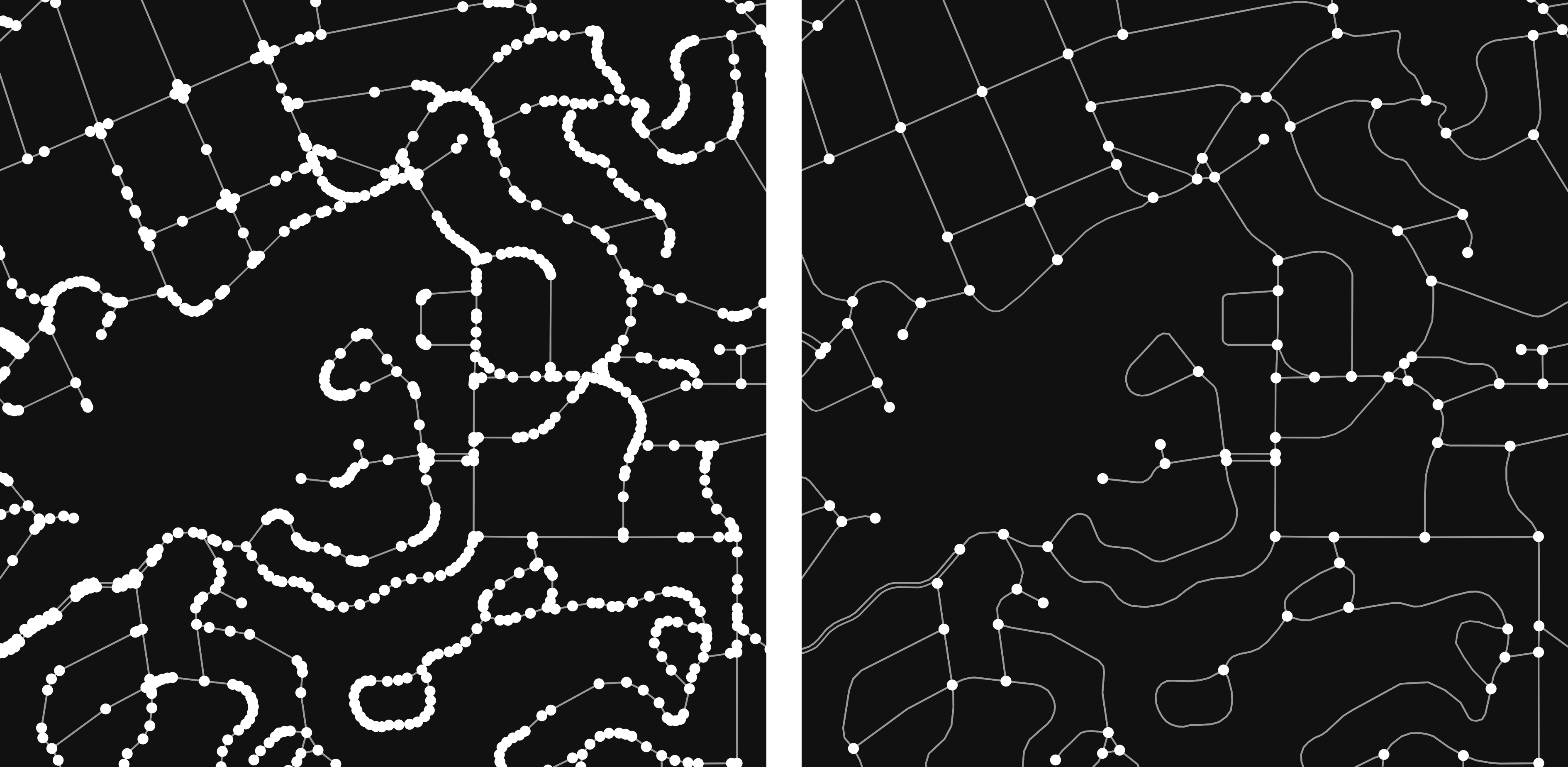}
    \caption{Graph of a town's street network before (left) and after (right) edge simplification. Source: \citet{boeing_graph_2025}.}\label{fig:graph_simplification}
\end{figure*}

\textit{Edge simplification}, put simply, merges adjacent edges for a better model. It cleans up the graph's topology so that nodes represent intersections or dead-ends and edges represent street segments, as illustrated by Figure~\ref{fig:graph_simplification}. This is important because in OpenStreetMap's raw data, ways comprise sets of straight-line segments between nodes: that is, nodes are also vertices for streets' curving line geometries, not just intersections and dead-ends. By default, OSMnx simplifies this topology by deleting non-intersection/dead-end nodes, merging the edges between them into a new \enquote{simplified} edge, and retaining the complete, true edge geometry as an edge attribute. When multiple OpenStreetMap ways are thus merged into a new graph edge, those ways' attribute values can be aggregated into a single value.

\textit{Node consolidation}, put simply, merges nearby nodes for a better model. This is important because many real-world street networks feature complex intersections and traffic circles, resulting in a cluster of graph nodes where there is really just one true intersection as considered in transport planning or urban design. Similarly, divided roads are often represented by separate centerline edges. The intersection of 2 divided roads thus creates 4 nodes where each edge intersects a perpendicular edge---but these 4 nodes represent a single intersection in the real world. OSMnx can consolidate such complex intersections into a single node and optionally rebuild the graph's edge topology accordingly. When multiple OpenStreetMap nodes are thus merged into a new graph node, the nodes' attribute values can be aggregated into a single value.

Edge simplification and node consolidation offer several benefits. They produce a more accurate model that better represents the real world. This in turn yields more accurate network measures, for example by not overcounting complex intersections when calculating intersection density or by not underrepresenting street segment lengths. Finally, many graph algorithms' time complexity scales with node or edge count. By generating a graph with (often drastically) fewer nodes and edges---yet no loss of accuracy---many algorithms will complete much faster. This matters most when analyzing large urban networks where runtime becomes a major constraint.

\subsubsection{Converting, Projecting, and Saving}

OSMnx's \texttt{convert} module can convert a NetworkX \texttt{MultiDiGraph} model to a NetworkX \texttt{MultiGraph} if the user prefers an undirected representation of the network for specific analytical purposes, as discussed in Section~\ref{sec:street_network_analysis}. It can also convert to a NetworkX \texttt{DiGraph} if the user prefers a directed graph without any parallel edges. OSMnx can also convert a \texttt{MultiDiGraph} to and from node and edge GeoPandas \texttt{GeoDataFrame} objects. The resulting nodes \texttt{GeoDataFrame} is indexed by OpenStreetMap node ID, and the resulting edges \texttt{GeoDataFrame} is multi-indexed by endpoint node $u$, endpoint node $v$, and a \textit{key} (to differentiate parallel edges), just as a \texttt{MultiDiGraph} edge is identified by its $u$, $v$, \textit{key} ordered triplet. This also allows users to load arbitrary node and edge ShapeFiles or GeoPackage layers as \texttt{GeoDataFrame}s then convert them to a \texttt{MultiDiGraph} for network analysis.

As these models are all spatial graphs, they have coordinate reference system (CRS) metadata. OpenStreetMap's default CRS is EPSG:4326, but users can project a graph to any other CRS using the \texttt{projection} module. If a user is unsure which CRS to project to, OSMnx can automatically determine an appropriate Universal Transverse Mercator CRS for the operation, based on the graph nodes' centroid.

Finally, using the \texttt{io} module, users can save a graph to disk as a GraphML file (to load into other network analysis software, including OSMnx), a GeoPackage (to load into other GIS software), or an OSM XML file (the standard OpenStreetMap data interchange format).

\subsubsection{Elevation}

Topography is essential to understanding street network form, but street network analyses too often ignore elevation as it can be difficult to acquire and attach to a graph model \citep{boeing_street_2022}. OSMnx's \texttt{elevation} module lets users automatically add elevation attributes to a spatial graph's nodes from either a local raster file or the Google Maps Elevation API.\@ Once all nodes have elevation attributes, users can automatically calculate edge grades (i.e., rise-over-run inclines), analyze the steepness of certain streets, or use elevation change in an impedance function for routing, as discussed below.

\subsubsection{Map Matching and Routing}

OSMnx offers basic map matching and routing functionality. The \texttt{distance} module can match a list of coordinates to each's nearest node or edge using a fast spatial index and vectorized operation. This can be useful for converting an origin-destination matrix of coordinates (such as geocoded addresses) into corresponding nearest nodes for path solving.

The \texttt{routing} module can solve shortest paths for network routing---parallelized with multiprocessing---using different edge weights (e.g., distance, travel time, elevation change, etc.). OpenStreetMap has a \enquote{maxspeed} tag representing streets' maximum speed limits, but it tends to be sparse for most cities. To address this problem, the \texttt{routing} module can impute missing edge maximum speeds based on observed values across other edges of the same type in the graph. Such imputation can be imprecise, but the user can override it by passing per-type local speed limits. Once all edges have maximum speed attributes, the module can also automatically calculate free-flow traversal times for each edge.

\subsubsection{Spatial Network Measures}

The \texttt{stats} module can calculate a variety of geometric and topological network measures \citep{boeing_street_2022}. These measures define streets as the edges in an undirected representation of the graph to prevent double-counting the bidirectional edges of a two-way street. Users can automatically calculate common measures from transport planning, urban design, and network science, including intersection density, circuity, average node degree, centrality, and many others. Users can also use NetworkX directly to calculate dozens of additional topological network measures. OSMnx's \texttt{bearing} module can calculate the streets' compass bearings and the network's orientation entropy \citep{boeing_urban_2019}. Finally, the \texttt{utils\_geo} module exposes a set of geospatial utility functions for graph-constrained spatial sampling, point interpolation, and more.

\begin{figure*}[tb]
    \centering
    \includegraphics[width=1\textwidth]{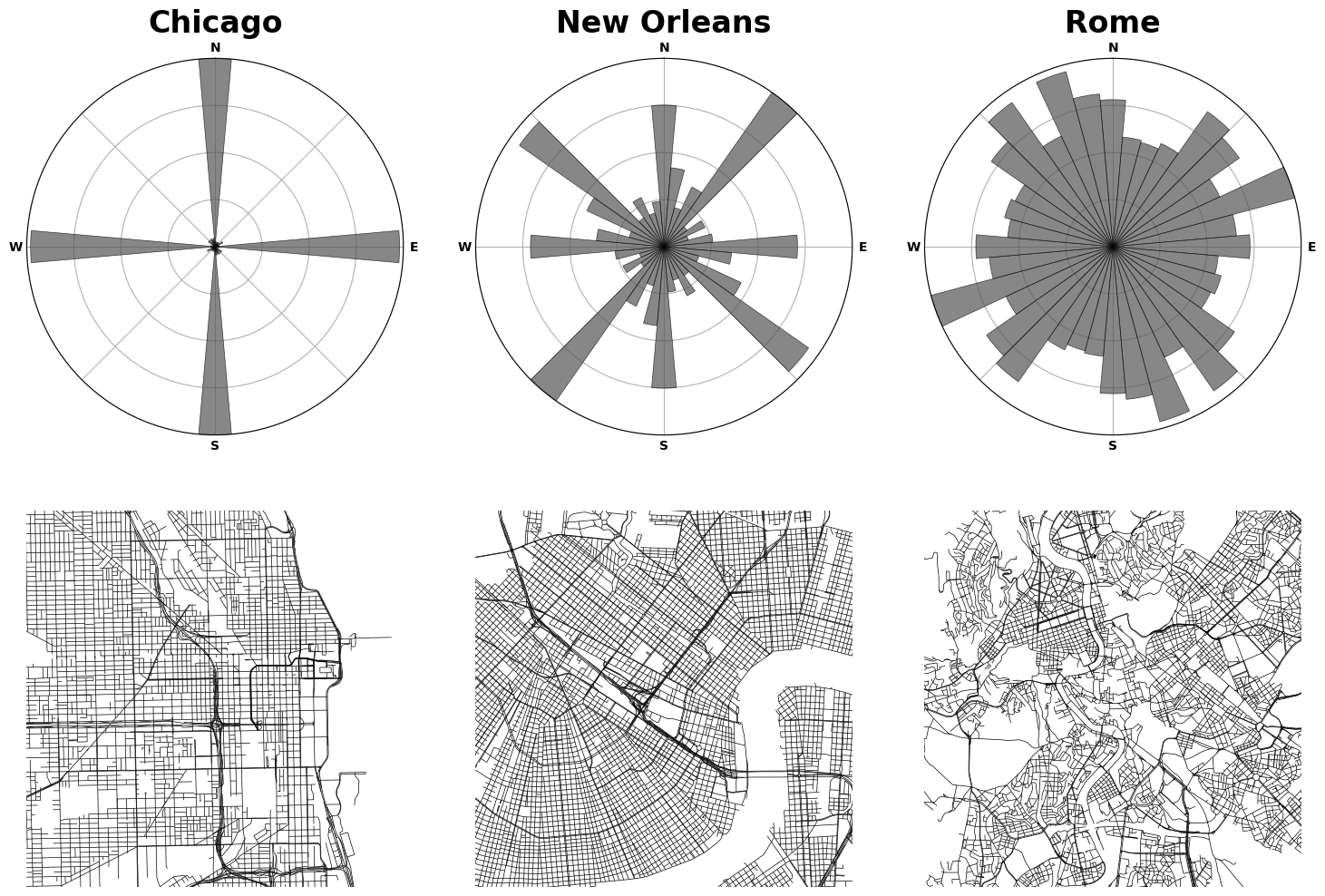}
    \caption{Polar histograms and corresponding street maps illustrating orientation entropy: low (Chicago), medium (New Orleans), and high (Rome). Histogram bar directions represent street compass bearings and lengths represent relative frequency. Source: \citet{barthelemy_review_2024}.}\label{fig:street_orientations}
\end{figure*}

\begin{figure*}[tb]
\centering
\includegraphics[width=1\textwidth]{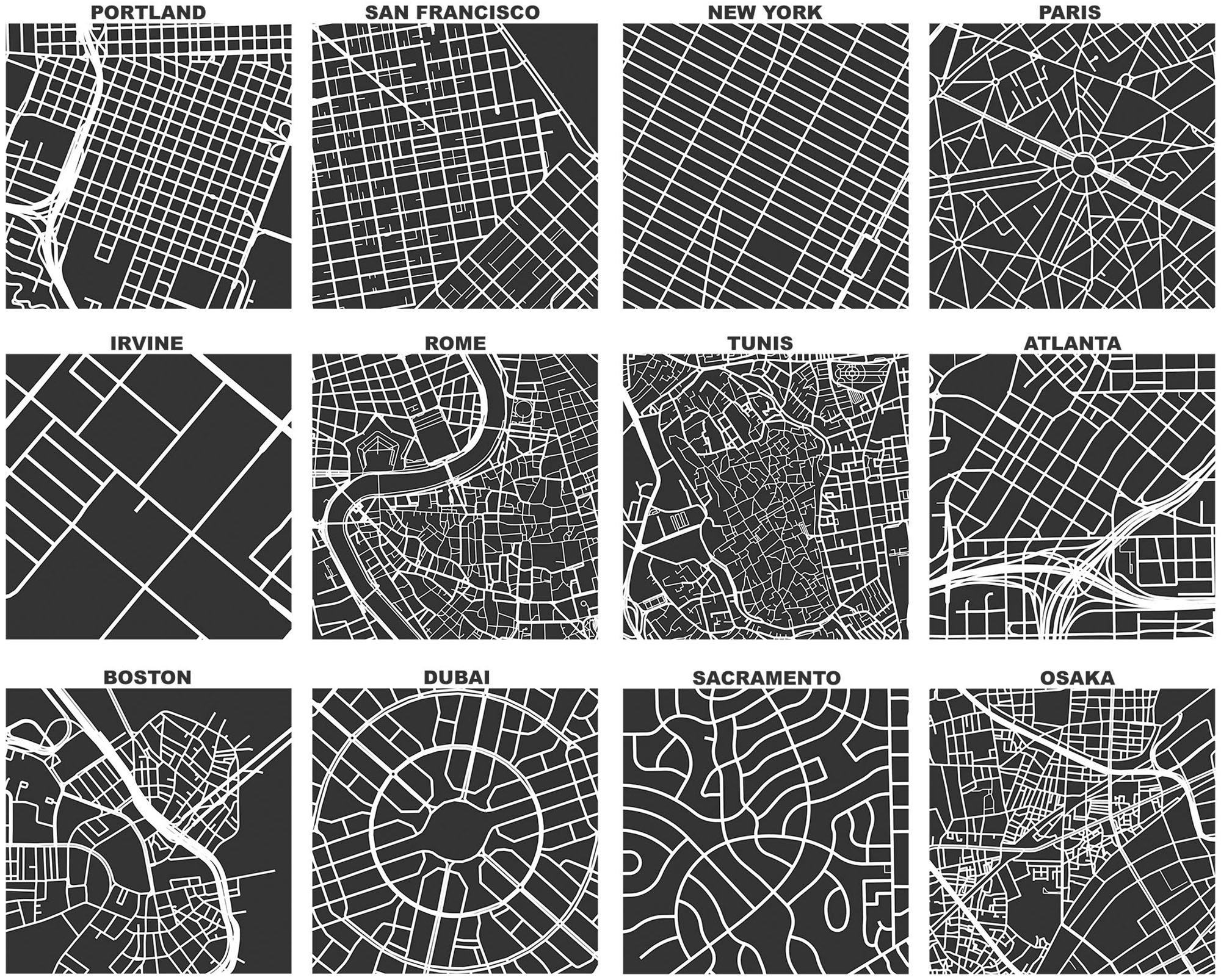}
\caption{Figure-ground diagrams of one square mile of each city, allowing us to compare street network form in different places. Source: \citet{boeing_spatial_2021}.}\label{fig:figure_ground}
\end{figure*}

\subsubsection{Visualization}

\setcitestyle{square}
Using OSMnx's \texttt{plot} module, users can visualize a wide variety of network characteristics, routes, building footprints, network orientation polar histograms (illustrated by Figure~\ref{fig:street_orientations} and discussed in \citet{boeing_urban_2019}), and network figure-ground diagrams (illustrated by Figure~\ref{fig:figure_ground} and discussed in \citet{boeing_spatial_2021}).Users can also easily explore networks, routes, or urban amenities as interactive web maps.
\setcitestyle{round}

\subsection{Installation and Configuration}

The OSMnx project's source code\endnote{The OSMnx source code is hosted at \href{https://github.com/gboeing/osmnx}{https://github.com/gboeing/osmnx}} is publicly hosted on GitHub. Users can install the package from the PyPI package repository or from the Anaconda package repository, as detailed in the package documentation's\endnote{The OSMnx documentation is available at \href{https://osmnx.readthedocs.io/}{https://osmnx.readthedocs.io/}} installation instructions. To get started, users can read the documentation's Getting Started guide then work through the OSMnx Examples Gallery\endnote{The OSMnx Examples Gallery is hosted at \href{https://github.com/gboeing/osmnx-examples}{https://github.com/gboeing/osmnx-examples}} for step-by-step tutorials, feature demonstrations, and sample code.

Once installed and imported, users can configure OSMnx through its \texttt{settings} module. Here they can adjust logging behavior, server response caching, server endpoints (including pointing to locally hosted instances), and more. Users can also configure OSMnx to retrieve historical snapshots of OpenStreetMap data as of a certain date.

\section{Lessons Learned in Geospatial Software}

Even \textit{if} OSMnx is good software today, it certainly was not always. The package began life in 2015 and its first version 0 beta release in early 2016 was a single monolithic module that geocoded place queries, downloaded raw OpenStreetMap data, built graph models, and projected graphs to different coordinate reference systems. It was ugly, brittle, and impenetrable---but it worked to pipeline OpenStreetMap data into graph algorithms and analyses in tools like NetworkX or IGraph. The following couple of years saw development continue on two primary fronts: first, in support of specific research needs or grants, with new functions or modules added accordingly. Second, as its user base grew, the project became more community-driven with users submitting bug reports or feature requests and contributing fixes or enhancements to the code base.

This culminated in a major milestone in 2021 with the release of OSMnx version 1.0. In particular, this included several new features and a massive refactoring that entailed some API-breaking changes deemed worthwhile as its new modular design made the toolkit more comprehensible for new users. Whereas functionality was previously lumped together in poorly architected modules and a monolithic namespace, the new API's design organized functions into thematic (and thus discoverable and predictable) modules.

OSMnx finally reached version 2.0 at the end of 2024, with several new features in addition to under-the-hood best practices aimed at making the package more user-friendly and sustainable for years to come. It is worth considering lessons learned---and suggestions for other researchers' geospatial software projects---in API design, development pipelines, dependency ecosystems, and the future of open science.

\subsection{API Design}

A software package's API is the interface through which users interact with it. It allows the package to conceal its internals while exposing a stable, well-documented set of functions and parameters. In Python and other languages, these functions are often organized into modules of related functionality.

The OSMnx v2.0 API exposes 17 public modules, as detailed in Section~\ref{sec:modular_organization}. Over years of development, the project has stabilized this API around several principles. It aims for parsimony\endnote{The principle of parsimonious design is perhaps best expressed by Antoine de Saint-Exupéry's famous aphorism: \enquote{Perfection is finally attained not when there is no longer anything to add, but when there is no longer anything to take away.}} and clarity: functionality is exposed via the fewest necessary functions controlled by the fewest necessary parameters to provide the user an approachable streamlined experience. Such API design is difficult to achieve---especially upfront and especially for accretionary academic projects---because it requires considerable foresight and software design skill, both of which academics usually only develop along the way. But these design and streamlining efforts in recent years have supported the long-term maintainability of OSMnx's codebase and promoted the DRY (don't repeat yourself) principle of software design.

Good documentation is also key to making an API usable and comprehensible. The OSMnx project emphasizes complete and clear documentation including a \enquote{getting started} guide for newcomers and a comprehensive examples gallery containing tutorials, feature demonstrations, and sample code. Every module and function contains a \textit{docstring} explaining its purpose, important notes or the scientific basis of algorithms and implementations, and the details of parameters and return values. These docstrings make the code \enquote{self-documenting} in that their consistent format and contents allow an up-to-date API reference to be automatically rendered and published on the web every time the codebase's main branch is updated.

Lastly, OSMnx is fully type-annotated. This seemingly inconsequential observation deserves some explanation. One of Python's benefits is that it is a dynamically-typed language: its objects and variables do not have their types specified by the coder. Rather, a variable implicitly takes on the data type of whatever value it is assigned, each time it is assigned. The drawback is that bugs can be hard to detect when a coding mistake allows a variable to be assigned a value with an unexpected data type, which then propagates through the system or causes an inscrutable error for edge cases. Type annotation allows the coder to specify objects' types without enforcing them. This adopts the benefits of type signatures in statically-typed languages like C or Rust by providing (1) type hints to the user and (2) the option of enforcing static typing in the software development pipeline, as discussed below.

\subsection{Development Pipeline}

The term \enquote{DevOps} refers to integrating software development with deployment and operations processes. OSMnx does this by automating its testing, packaging, and deployment workflows. This includes continuous integration: the practice of maintaining a clean main code branch and frequently merging code changes after they have undergone code review and passed a series of tests. Automating these mundane processes is particularly important for relatively small and resource-constrained open-source projects.

The OSMnx project has converged around a modern set of Python-oriented tools to do this. It uses Git for version control and GitHub for project management and collaboration, and its continuous integration test workflow includes (1) a large suite of unit tests run via pytest, (2) code coverage tests via Codecov to ensure that newly added code is fully tested (OSMnx's current codebase coverage exceeds 98\%), (3) code quality, linting, and formatting tests via Ruff to ensure that coding best practices and project standards are followed, (4) static type checking via mypy and runtime type checking via Typeguard to ensure consistent typing throughout the package, and (5) documentation linting and building via numpydoc and Sphinx to ensure that docstrings meet standards and that the documentation website is always up to date. Some of these standards were decided on and enforced during v2.0 development (requiring significant changes), but it would be much easier to build around them upfront at the start of a project.

Other workflows run on a schedule to test against both minimum required dependency versions (for backwards compatibility) and pre-release dependency versions (for future compatibility). Finally, two other workflows handle continuous deployment whenever a new release is tagged on GitHub. One builds a Docker image and publishes it to DockerHub, and the other builds the Python package for distribution and publishes it on PyPI for users to install.

\subsection{Dependency Ecosystems}

At its inception, the OSMnx project chose the Python programming language because it is general-purpose, easy to learn (which reduces barriers to entry for newcomers), and offers a large ecosystem of existing packages providing data structures and functionality. This also means that it can be incorporated into GIS-based workflows via software like ArcGIS and QGIS that use Python for extensibility. In some ways, OSMnx exists as \enquote{glue code} in that its models and analyses connect different scientific libraries with scientific data.

The development of OSMnx v2.0 focused on interoperability with the rest of the Python geospatial stack. It depends on six other packages: GeoPandas, NetworkX, NumPy, Pandas, Requests, and Shapely. For extended functionality, it optionally depends on Matplotlib, Rasterio, rio-vrt, SciPy, and scikit-learn. Separating these required versus optional dependencies allows users to install a lightweight minimal environment for core functionality, or a more extensive environment for complete functionality. The project also reduced its overall number of dependencies without reducing its functionality by streamlining its internals.

OSMnx is tightly coupled to two of its dependencies---GeoPandas (for spatial dataframes) and NetworkX (for graph data structures and algorithms)---as well as OpenStreetMap itself as a data source. Such coupling allows the project to avoid reinventing the wheel by instead offloading the implementation of basic data structures and algorithms to its dependencies. For example, it implements several spatial indices for search routines, including an R-tree (via Shapely), a \textit{k}-d tree (via Scipy), and a ball tree (via scikit-learn). Reusing existing data structures such as these saves development time and allows each project to specialize in its own contribution without having to innovate everything from scratch.

The future of this ecosystem seems promising. In particular, as an interpreted rather than compiled language, Python has always been hamstrung by relative slowness---but the language's latest releases have made significant speed improvements. Whereas it historically relied on compiled C/C++ extensions for faster execution, the past two years have seen Python increasingly serving as a wrapper around extremely fast Rust code libraries. OSMnx may benefit from this in the future through projects like RustworkX \citep{Treinish2022}, a reimplementation of NetworkX in Rust. Additional speed improvements can be found by shifting linear algebra-based graph algorithms from the CPU to the GPU, as with the cuGraph library. Finally, new data offer new possibilities. Most prominently, Overture Maps now provides worldwide building footprints, roads, political boundaries, addresses, and land cover data---all with a well-structured schema and validation checks. Though new, this initiative deserves geographers' attention.

\subsection{Geospatial Software for Open Science}

The academic geography community can advance open science by supporting the open software on which it relies. The first pathway is through educating the next generation on the benefits of open-source solutions and teaching the \enquote{whys} and \enquote{hows} of algorithms and implementations rather than relying on black-box point-and-click graphical user interfaces. The second pathway is through better incentives and funding. Academic promotion and tenure processes should recognize and reward the scholarly value of open-source research software and open data contributions to geography's empirical and theoretic endeavors. Additional funding opportunities for both building and maintaining open research software are key to realizing their positive externalities across communities of scholarship and practice. Finally, the third pathway is through academic publishing. Many GIScience journals are at the forefront of requiring submission of the data and code underlying quantitative manuscripts. This is essential for reproducible and interpretable science, but is not enough. We also need our tools to be \textit{reusable}---that is, well-designed, well-documented, well-maintained, generalizable, accessible, and easy to use.

\section{Conclusion}

The open science movement calls for open, reusable research tools to support better science. The OSMnx project contributes to these goals by developing and maintaining such a tool for modeling and analyzing urban networks and amenities anywhere in the world from OpenStreetMap data. It embodies the relevant domain theory from urban planning, network science, and geographical information science in a tool designed to \enquote{just work.} The OSMnx project improves reproducibility by allowing users to automatically model networks with clearly defined spatial extents, thoroughly documented modeling decisions, and theoretically-sophisticated results. For example, its graph models are primal, nonplanar, directed, weighted, multigraphs with possible self-loops to capture the diversity of network types and characteristics around the world. Its graph creation methods attenuate artificial perimeter effects, and its graph simplification methods generate models that flexibly match real-world expectations and use. In turn, researchers need not reinvent the wheel or make dozens of ad hoc basic modeling decisions for each urban network analysis project.

This article documented the OSMnx package's modern organization, capabilities, and theoretical underpinnings. It offered some lessons learned over ten years of geospatial software development and some suggestions for others working on similar projects. Better tools, led by scholarly experts, represent one of the key paths forward toward better and more open science. Contributing to this shared tools commons benefits everyone across disciplines.

\section*{Acknowledgments}

The author wishes to thank the developers and maintainers of the many other packages on which the OSMnx project depends, as well as the contributors who have lent their time and expertise in proposing features, solving bugs, and contributing code to the OSMnx code base over the years. Figure~\ref{fig:graph_simplification} is reprinted from \citet{boeing_graph_2025} under the terms of the CC BY-NC-ND 4.0 international license. Figure~\ref{fig:street_orientations} is reprinted from \citet{barthelemy_review_2024} under the terms of the CC BY 4.0 international license. Figure~\ref{fig:figure_ground} is reprinted from \citet{boeing_spatial_2021} with permission from Elsevier.

\IfFileExists{\jobname.ent}{\theendnotes}{}

\setlength{\bibsep}{0.00cm plus 0.05cm} 
\bibliographystyle{apalike}
\bibliography{references}

\end{document}